# Bennu's global surface and two candidate sample sites characterized by spectral clustering of OSIRIS-REx multispectral images

J. L. Rizos[a,b], J. de León[a,b], J. Licandro[a,b], D. R. Golish[c], H. Campins[d], E. Tatsumi[a,b,e], M. Popescu[f], D. N. DellaGiustina[c], M. Pajola[g], J.-Y. Li[h], K. J. Becker[c], and D. S. Lauretta[c]

a Instituto de Astrofísica de Canarias, C/Vía Láctea s/n, E-38205 La Laguna, Tenerife, Spain

b Departamento de Astrofísica, Universidad de La Laguna, E-38206 La Laguna, Tenerife, Spain

c Lunar and Planetary Laboratory, University of Arizona, 1415 N. Sixth Ave., Tucson, AZ 85705-0500, USA

d Department of Physics, University of Central Florida, P.O. Box 162385, Orlando, FL 32816-2385, USA

e Department of Earth and Planetary Science, University of Tokyo, 7-5-1 Hongo, Bunkyo-ku, Tokyo, Japan

f Astronomical Institute of the Romanian Academy, 5 Cuțitul de Argint, 040557 Bucharest, Romania

g INAF – Astronomical Observatory of Padova, Vic. Osservatorio 5, 35122 Padova, Italy

h Planetary Science Institute, Tucson, AZ 85719, USA

## ABSTRACT

The OSIRIS-REx spacecraft encountered the asteroid (101955) Bennu on December 3, 2018, and has since acquired extensive data from the payload of scientific instruments on board. In 2019, the OSIRIS-REx team selected primary and backup sample collection sites, called Nightingale and Osprey, respectively. On October 20, 2020, OSIRIS-REx successfully collected material from Nightingale. In this work, we apply an unsupervised machine learning classification through the K-Means algorithm to spectrophotometrically characterize the surface of Bennu, and in particular Nightingale and Osprey. We first analyze a global mosaic of Bennu, from which we find four clusters scattered across the surface, reduced to three when we normalize the images at 550 nm. The three spectral clusters are associated with boulders and show significant differences in spectral slope and UV value. We do not see evidence of latitudinal non-uniformity, which suggests that Bennu's surface is well-mixed. In our higher-resolution analysis of the primary and backup sample sites, we find three representative normalized clusters, confirming an inverse correlation between reflectance and spectral slope (the darkest areas being the reddest ones) and between $b'$ normalized reflectance and slope. Nightingale and Osprey are redder than the global surface of Bennu by more than 1σ from average, consistent with previous findings, with Nightingale being the reddest ($S' = (-0.3 \pm 1.0) \times 10^{-3}$ percent per thousand angstroms). We see hints of a weak absorption band at 550 nm at the candidate sample sites and globally, which lends support to the proposed presence of magnetite on Bennu.

## 1. Introduction

The primary goal of NASA's Origins, Spectral Interpretation, Resource Identification, and Security–Regolith Explorer (OSIRIS-REx) mission is to return pristine carbonaceous regolith from the surface of near-Earth asteroid (101955) Bennu (Lauretta et al., 2017). Bennu, as a B-type primitive asteroid (Clark et al., 2011), likely harbors information about the early stages of our Solar System. Bennu is a ~500-m-diameter rubble pile (e.g., Barnouin et al., 2019), formed from the reaccumulated fragments of a larger, catastrophically disrupted parent body (Michel and Ballouz et al. 2020). It has a globally rough surface showing a wide range of particle size and albedo characteristics (Lauretta et al., 2019; DellaGiustina et al., 2019). Its average visible-near-infrared spectral slope is moderately blue (negative) (DellaGiustina et al., 2020; Clark et al., 2011). Color and reflectance variation suggest primordial heterogeneity —inherited from its parent body and other asteroids (DellaGiustina et al., 2021)— and different exposure ages (DellaGiustina et al., 2020).

The OSIRIS-REx spacecraft carries a camera suite, OCAMS, consisting of three cameras: PolyCam, MapCam, and SamCam (Rizk et al., 2018; Lauretta et al., 2017). For this work, we focus on multispectral data collected by the MapCam medium-field-of-view imager. MapCam has a 125-mm focal length and a focal ratio of f/3.3, which provides a ~ 4° field of view (Rizk et al., 2018). It has four passbands filters based on the Eight-Color Asteroid Survey (ECAS, Tedesco et al., 1982): $b'$, $v$, $w$, and $x$, with effective wavelengths at 473, 550, 698, and 847 nm, respectively. The original ECAS $b$ filter was shifted toward longer wavelengths to improve optical and radiometric performance and to minimize aging effects due to radiation (Rizk et al., 2018). Compared to other spacecraft imagers, MapCam provides color information with remarkably high spatial resolution. The MapCam bands were chosen to permit the detection and characterization of an absorption band centered at 700 nm, if present. This band is associated with the presence of Fe-bearing phyllosilicates—i.e., silicates that have been altered by liquid water—and strongly correlates with the well-known absorption feature around 3000 nm, indicative of hydration (Rivkin et al., 2015, Vilas, 1994). These filters also make it possible to measure the spectral slope, which can be used to identify space weathering effects (DellaGiustina et al., 2020).

The OSIRIS-REx spacecraft encountered Bennu on December 3, 2018 and has since acquired extensive data from OCAMS and other scientific, navigation, and communication instruments on board. The global observational phases of the mission took place in 2019, including Detailed Survey (DS), where the spacecraft acquired images of Bennu's surface in all MapCam color filters in rapid succession to capture the same scene under near-uniform photometric conditions. The asteroid was observed at 12:30 p.m. local time (7.5° phase value) to provide low- incidence and emission angles and the spacecraft slowly slewed to provide as much filter to-filter image overlap as possible for co-registration (DellaGiustina et al., 2018). In particular, DS Baseball Diamond Flyby 2B took place September 26, 2019, and was used to build global mosaics in each MapCam filter. These data were used to study various regions of interest as candidate sites for sample collection. Later, during the Reconnaissance A phase (Recon A) in October 2019, the spacecraft left a terminator plane orbit and approached the surface to take high-resolution MapCam color images of the final candidate sites under optimum photometric conditions (DellaGiustina et al., 2018). In December 2019, the OSIRIS-REx team selected Nightingale (56.0° N, 42.1° E) and Osprey (11.7° N, 86.5° E) as the primary and backup sample collection sites, respectively. Finally, on October 20, 2020, OSIRIS-REx successfully collected a sample from Nightingale.

The goal of this work is to spectrophotometrically characterize the local-scale primary and backup sample collection sites in context with the global surface of Bennu via a clustering approach. We aim to shed light on how the local-scale (candidate sample site) surface properties resemble or differ from those observe at the global scale. To fulfill our purpose, we apply the methodology described and validated Rizos et al. (2019) to the MapCam images in the global color mosaic of Bennu and those collected at Nightingale and Osprey separately. Our methodology consists of an unsupervised machine learning classification through the K-Means algorithm, which allows us to identify clusters with spectral similarities. In Section 2, we explain the data processing and the methodology used. In Section 3, we present the results of our global and local (sample site) characterization. Section 4 discusses and summarizes our findings.

## 2. Data analysis

### 2.1. Data preparation

The global mosaic was built using MapCam images acquired during Flyby 2B (DS phase) at an altitude of 3.8 km and with a pixel scale of 25 cm. For the primary and backup sample collection sites, in addition to those images acquired during the Flyby 2B, we also use images from Recon A at an altitude of 1.202 and 1.142 km for Nightingale and Osprey, respectively, and with a pixel scale of 6 cm. All MapCam images were calibrated in radiance factor (RADF[1]) following the procedure described in Golish et al. (2020). The calibration pipeline can be summarized as bias/dark subtraction, charge smear correction, flat field normalization and radiometric conversion. After radiometric calibration, it is necessary to apply photometric corrections to standardize phase ($\alpha$), emission ($e$), and incidence ($i$) angles for each pixel. An in-depth photometric study was carried out in Golish et al., (2021), which concluded that a ROLO (Robotic Lunar Observatory) model presents the smallest $\chi^2$ error (see their Table 3) and is thus best suited for Bennu's surface. This model assumes that reflectance can be expressed as a product of a phase function, $A(\alpha)$, and a disk function, $d(i, e)$, as shown in Eqs. (1) – ((3). We therefore chose this model for correcting photometrically MapCam images to $(\alpha, i, e)$ = (0º, 0º, 0º) as reference angles[2].

$$Reflectance = A(\alpha) \cdot d(i, e) \quad (1)$$

$$A(\alpha) = C_0 e^{-C_1 \alpha} + A_0 + A_1 \alpha + A_2 \alpha^2 + A_3 \alpha^3 + A_4 \alpha^4 \quad (2)$$

$$d(e, i) = \frac{\cos(i)}{\cos(i) + \cos(e)} \quad (3)$$

Phase, emission, and incidence angle backplanes (necessary for photometric corrections), as well as image registration, projection, and mosaicking, are obtained using a version of ISIS3[3] modified to support processing with tessellated 3D shape models. The procedure to create the mosaics is detailed in DellaGiustina et al. (2020). For the global mosaic analysis, we used the v28 global shape model constructed using stereophotoclinometric techniques with a ground sample distance (GSD) of 80 cm, whereas for Nightingale and Osprey, we used OLA-based tile shape models with a GSD of 5 cm (Barnouin et al., 2020) derived from data collected by the OSIRIS-REx Laser Altimeter (Daly et al., 2017). We smoothed the emission and incidence backplanes by running a lowpass filter through an average boxcar of 21×21 pixels to avoids artifacts caused by corners or edges of the boulders. Photometric models have limited accuracy at incidence or emission angles > 80°; hence, we excluded all pixels acquired at these conditions from the data used in this study.

---

[1] For consistency, in this paper we use reflectance and RADF interchangeably, according to the criterion followed by other authors in analogous works (Schröder et al., 2017).

[2] The most used reference geometries for corrections are (α, i, e) = (0º, 0º, 0º), which corresponds to the normal reflectance, and (α, i, e) = (30º, 30º, 0º), which is a common laboratory setting. We test both cases and our clustering results do not change at all; corrections affect all pixels equally.

[3] ISIS3 is a robust tool to manipulate imagery collected by NASA planetary missions (https://isis.astrogeology.usgs.gov).

Another source of inaccuracies are MapCam pixels near saturation, with digital numbers (DN) above the linearity range of the CCD (>14000 DN). Thus, we also removed those pixels. Finally, shadows and other areas not directly illuminated in MapCam images have a low signal-to-noise ratio (SNR). To do this, we carry out a direct measurement on the images, verifying that regions with reflectance values lower than 0.027 in the *v*-filter correspond to areas without direct lighting. Hence, we removed these regions before analysis.

## 2.2. Clustering analysis

Using calibrated and photometrically corrected images, we applied the K-Means algorithm (MacQueen, 1967), which is an unsupervised partitioning technique. This method is commonly used in different fields of astronomy to identify clusters when no previous knowledge of a data set exits. In the color mosaics of Bennu, each pixel has four values corresponding to the reflectance of each MapCam band. In four-dimensional space (spectral space), each pixel is represented by a point with four coordinates, and the totality of the pixels form a point cloud. The clustering technique identifies different subgroups within this cloud. If we assume that there are n representative subgroups, our partitioning algorithm K-Means generates n random points (called centroid seeds) in our four-dimensional space as representative clusters. Then, the algorithm calculates the distance of each point to each of the representative clusters and, depending on the result, assigns membership to one group or another. After that, it calculates the centers of each data set, then computes the distances of each point to these new representative clusters and iterates this process consecutively to refine the result. The result of this procedure is the identification of many representative clusters (Rizos et al. 2019).

For this task, we used the scikit-learn Python package (Pedregosa et al., 2011). It allows us to run the K-Means algorithm using random centroid seeds, and the output is that which minimized the sum of the squared differences between each point and its center group. The package also permits customizing the convergence criteria and has a parallel computing option, making it an ideal tool for analyzing the millions of spectra included in Bennu's multispectral mosaic.

Considering only the shape of the obtained spectra (relative reflectance) can be even more useful and profitable than considering absolute reflectance. As demonstrated in Rizos et al. (2019), clustering with relative reflectance is more sensitive to shallow spectral features, such as the 700-nm absorption, as it excludes points with similar albedo but no 700-nm feature. Thus, we also applied the clustering analysis after normalizing the images by dividing the reflectance value of each filter by the reflectance at 550 nm (*v* filter).

To choose the optimal number of representative clusters, we employed a multiapproach criterion. First, we applied the Elbow method (Syakur et al., 2018), which computes the sum of squared errors (SSE) considering a range of the number of clusters and represents this quantity against the number of clusters. An elbow shape in this graphical representation points toward the suitable number. However, this elbow cannot be unambiguously identified (Ketchen and Shook 1996), especially in cases where the clusters do not strongly differ from each other. In those cases, the method only gives a range of possible solutions that must be inspected to determine the best one. In this work, we reproduced each solution within the wide-open elbow, increasing the number of clusters from two up to a given number for which the code begins to subdivide regions with spectral similarities into different clusters. From this point, the new clusters are within the error bars of the old ones, and the associated pixels of these newly identified clusters are randomly dispersed in the images, producing a loss of spatial coherence. Finally, to decide which of all these solutions is the best, we studied the correlation between reflectance, spectral slope, and UV behavior[4] ($b'$ filter) for each set of clusters.

*2.3. Characterization of spectral parameters*

In addition to identifying the number of clusters that best describes our data set, we defined several parameters to characterize MapCam spectra for comparison. The first parameter is the visible spectral slope, $S'$, which is computed with all reflectance bands by applying a linear fit using a least-squares method. To compute the error in the spectral slope, we use a Monte Carlo approach (De Prá et al., 2018), creating 1000 clones in each of the filter bands by using a Gaussian distribution centered at the central wavelength of the filter and with a width corresponding to the uncertainty associated with that measurement. The resulting spectral slope is the average value in units of percent per thousand angstroms (%/1000 Å), and the error is the standard deviation.

The other spectral parameters are the wavelength position of the center of the 700-nm band and its depth, $D_{700}$. To measure these, we first subtracted the continuum between the 550 nm ($v$ filter) and 847 nm ($x$ filter) wavelengths, then we fit a second-order polynomial to the reflectance at 550 nm ($v$ filter), 698 nm ($w$ filter), and 847 nm ($x$ filter). The wavelength position of the band center corresponds to the location of the minimum of the curve, whereas the band depth is computed as the difference, in percent, between a reflectance value of 1 and the reflectance at that minimum. The final values for the band center and depth and their corresponding errors are calculated using the same procedure as for the spectral slope. If band depth is above the 3σ level, we consider it as a positive detection of an absorption band. An analogue procedure is followed in cases where we detect a possible absorption band (UV upturn) at 550 nm ($D_{550}$), but using filters $b'$, $v$, and $w$ instead.

---

[4] Although the MapCam $b'$ filter does not collect ultraviolet photons properly, for consistency with other related works, we will continue calling UV to its measurements.

## 3. Results

### 3.1. Global mosaic

First, we characterize the global spectrophotometric behaviour of Bennu in the MapCam color images. Figure 1 shows the global mosaic used (*v*-filter), which covers a latitude range from aproximately -65º to +65º and >90% of the surface. The pixel scale of this mosaic is 25 cm, resulting in 14,233,185 pixels (each with a four-color spectrum) after removing unreliable data (see Section 2.1).

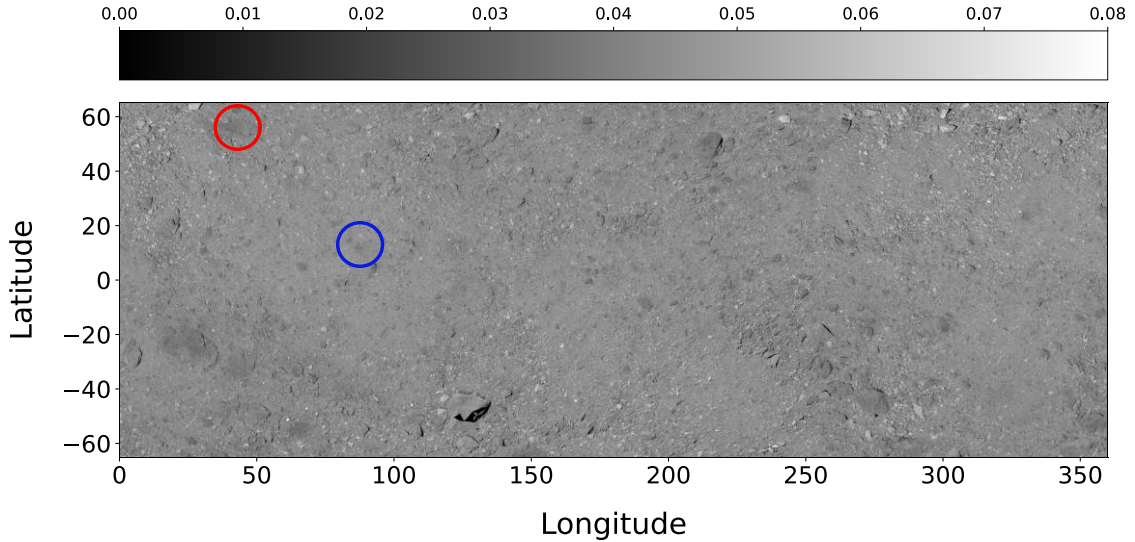

**Figure 1. The *v*-filter global mosaic of Bennu using Flyby 2B MapCam images, from DellaGiustina et al. (2020). Each image was photometrically corrected before assembly. The pixel scale of this mosaic is 25 cm, which makes more than 14 million pixels (spectra) after removing unreliable data. Circles mark Nightingale (red) and Osprey (blue), the OSIRIS-REx mission's primary and backup sample collection sites, respectively.**

To reduce noise, before running our clustering detection method, we applied a median boxcar, replacing each pixel value in the image with the median value of its neighbors. It is important to choose an appropriate size of the box, because if it is too large, it can blur the data and lose relevant information. So, we tested different boxcar sizes and we confirm that 7×7 pixels enhances the clusters without any loss of information in terms of global spectral distribution. Applying this boxcar could not allow the identification of small and unusual bright groups of isolated pixels, as those identified by DellaGiustina et al. (2021), but this specific task is outside the scope this work. For consistency, we use this median box henceforth in both the global and local site analyses.

We start with an absolute reflectance analysis. The Elbow method shows a shallow, wide-open elbow that extends up to eight/nine clusters (Figure 2a), so we analyze each possible solution in detail as explained in Section 2.2. Once we obtain the clusters for each possible solution, we normalize them at 550 nm to calculate spectral slope ($S'$), band depth and center at 550 and 700 nm ($D$), average reflectance at 550 nm ($\bar{r}$), and percentage of pixels. In all cases, we note the absolute reflectance value dominates this classification, given that the spectral shape for each cluster is practically identical within the uncertainty. Therefore, the difference between each solution is the number of divisions we make of the spectral continuum from the darkest to the brightest regions, where the lowest and highest albedo clusters are distributed around rockier terrains. Apart from absolute reflectance, the only difference manifests itself in a correspondence between low reflectance and redder spectral slope. We find four clusters as an intermediate suitable solution to characterize the global behavior of the surface without loss of information (Figure 2b,c). Except for the cluster C1, it is possible to measure an absorption band at 550 nm after subtracting the continuum. The drops are centered around 570 nm, but the absorption is not positive in any case according to our 3σ criterion (we indicate it with hyphens in Table 1). Regarding the 700-nm band, in none of the clusters is it possible to measure an absorption.

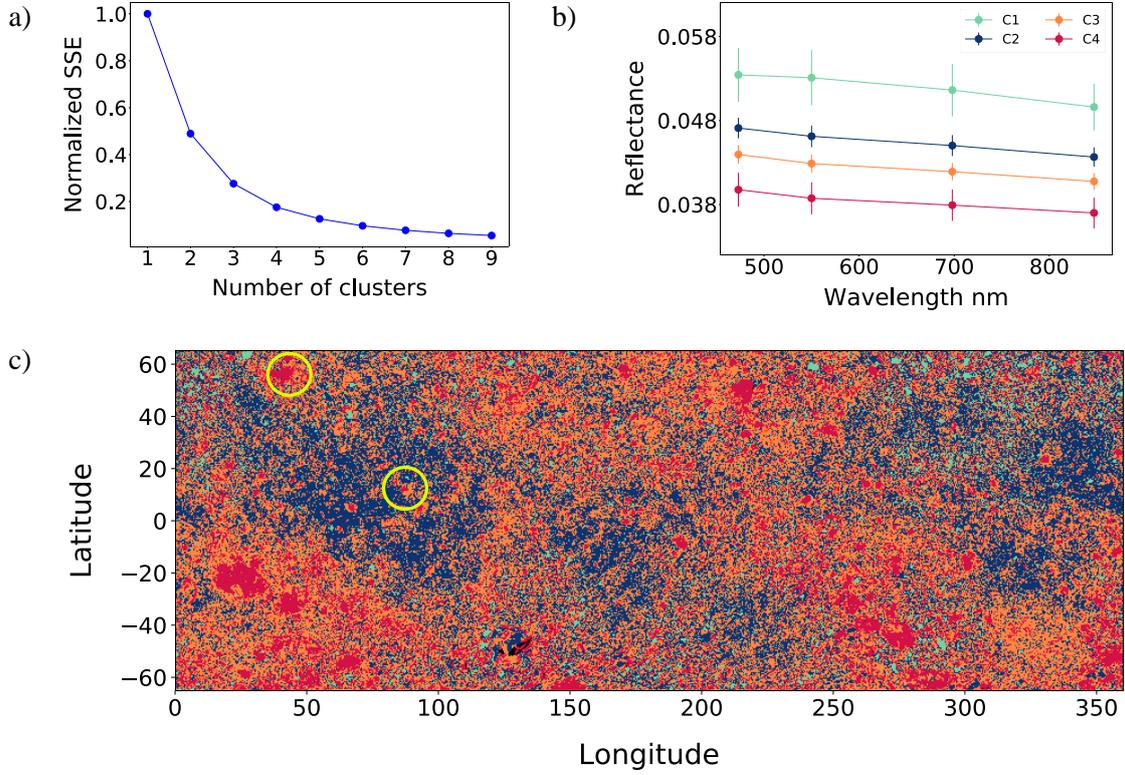

Figure 2. Outcomes from the application of the clustering technique to the global mosaic of Bennu. a) Normalized SSE vs. the number of clusters using the Elbow method for the global mosaic of Bennu in absolute reflectance. b) The four representative spectral clusters with error bars (standard deviation) are shown in different colors. c) The location of the identified clusters over the global map, using the same color code as in (a). Yellow circles mark Nightingale and Osprey. Black pixels are those that are out of the defined limits.

Table 1. Computed spectral slope ($S'$), band depth and center at 550 and 700 nm ($D$), average reflectance at 550 nm ($\bar{r}$), and percentage of occupied surface for each cluster (after correcting the latitudinal distortion caused by the projection) identified in the reflectance analysis of the global mosaic. Hyphens indicate cases where is not possible to measure an absorption band because there is no drop after subtracting the continuum.

| Cluster ID | C1 (Green) | C2 (Blue) | C3 (Orange) | C4 (Red) |
|---|---|---|---|---|
| $S'$ ($\times 10^{-3}$) (%/1000 Å) | -2.0 ± 1.9 | -1.9 ± 0.9 | -1.9 ± 0.8 | -1.8 ± 1.8 |
| $D_{550}$ (%) | — | 1.8 ± 2.3 | 2.0 ± 2.3 | 2.9 ± 3.9 |
| Center$_{550}$ (nm) | — | 570 ± 17 | 571 ± 17 | 569 ± 17 |
| $D_{700}$ (%) | — | — | — | — |
| Center$_{700}$ (nm) | — | — | — | — |
| $\bar{r}$ (550 nm) | 0.053 ± 0.003 | 0.046 ± 0.001 | 0.043 ± 0.001 | 0.039 ± 0.002 |
| Area (%) | 5.0 | 37.1 | 43.1 | 14.7 |

We next run our clustering analysis using image data normalized at 550 nm. Again, the Elbow method does not identify an obvious solution (Figure 3a), so we assessed all clustering solutions from two up to nine. Additionally, to support this assessment, we considered correlations between reflectance at 550 nm for each cluster before normalization, normalized $b'$ filter value, and spectral slope ($S_{vwx}$) obtained by applying a linear fit to normalized filters $v$, $w$ and $x$ using a least-squares method. For this solution, we find a Pearson correlation coefficient (PCC) of -0.99 between reflectance at 550 nm and $S_{vwx}$ and 0.91 between the normalized $b'$ filter value and $S_{vwx}$ (Figure 4). This indicates that more positive (redder) spectral slopes are correlated with low reflectance and high UV values. Thus, we conclude that three clusters summarize the normalized global spectral behavior of Bennu (Figure 3b,c).

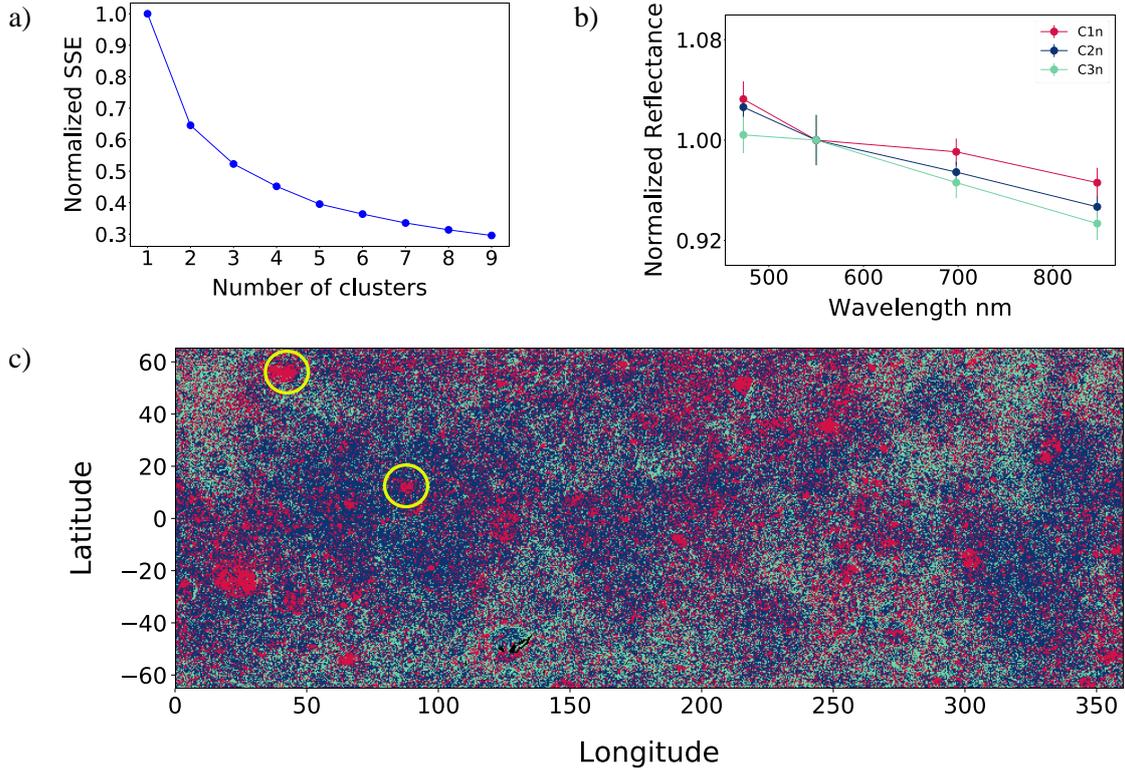

**Figure 3. Results from the application of the clustering technique to the normalized global mosaic of Bennu. a) Normalized SSE vs. the number of clusters using the Elbow method for the global mosaic of Bennu in relative reflectance. We see a wide-open, shallow "elbow" from two up to nine clusters. After analysis of all these possible solutions, we found an optimal number of three clusters as a number for characterizing the spectral behavior of the surface. b) The three representative spectral clusters with the corresponding error bars (standard deviation) are labeled with different colors. c) The location of the identified clusters over the analyzed area, using the same color code as in (a). Yellow circles mark Nightingale and Osprey. Black pixels are those that are out of the defined limits.**

The C2n cluster occupies more than 50% of the terrain and has a spectrum similar to ground-based spectral measurements of Bennu (de León et al., 2018). Global visible–near-IR spectra obtained by the OSIRIS-REx Visible and InfraRed Spectrometer (OVIRS; Reuter et al., 2018) also show a bluish spectrum (Simon et al., 2020a; Hamilton et al., 2019), as well as possible minor absorption bands centered at 550, 1050, 1400, 1800, and 2300 nm (Simon et al., 2020b). The only one of these features that is measurable by MapCam is the 550 nm absorption band, and we also observe it in our widely distributed C2n cluster. It presents a (1.4 ± 1.6) % absorption band, although the drop is below the associated uncertainty.

The C1n cluster (26.6%) is located in lower-reflectance regions, and the C3n cluster (21.0%) in the brightest areas. The clusters show a correlation between reflectance and spectral slope: darker areas have a more positive (redder) slope. A shallow band depth at 550 nm or a UV upturn appears for the C1n and C2n clusters. However, this does not satisfy our 3σ criterion when considering the error (Table 2). At 700 nm, only the C3n cluster shows a possible drop, but the uncertainty is larger than the depth.

Figure 5 shows a distribution of the percentage of pixels as a function of latitude. For this analysis, we divided the global mosaic (Figure 3) into 15 horizontal bands, each one occupying 8.68°. For each of these intervals we calculate the relative percentage of pixels, avoiding the latitudinal distortion caused by the projection. We see a decrease of the C3n cluster correlated with an increase of cluster C2n towards equatorial regions. Cluster C1n increase slightly towards northern regions.

In this global analysis, the Nightingale and Osprey sites, which are inside small craters, are redder than their neighboring terrain. Both sites fall into the C1n cluster of the global analysis. Other areas also show this redder spectral behavior, including the largest boulder found in Bennu, Roc Saxum (24° S, 28° E).

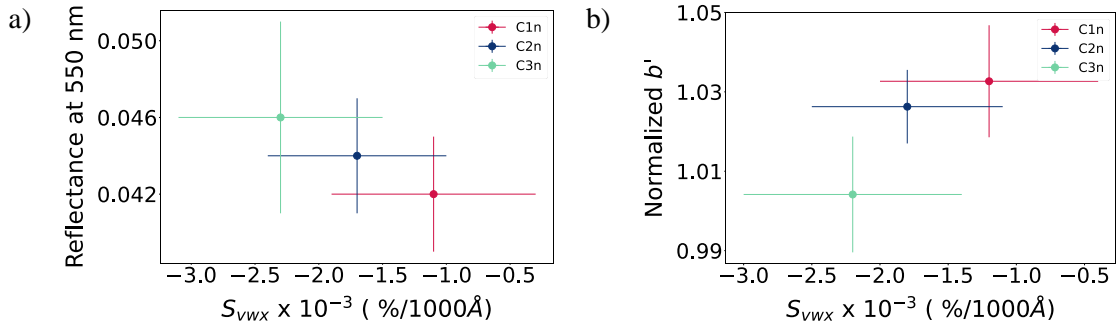

**Figure 4.** Reflectance at 550 nm before normalization (a) and normalized $b'$ filter value (b), against spectral slope ($S_{vwx}$) computed by applying a least-squares fit to filters $v$, $w$, and $x$. A correlation between reflectance and spectral slope is evident with a PCC of –0.99. Spectral slope and UV normalized value are correlated with a PCC of 0.91.

**Table 2.** Same as Table 1, but for normalized reflectance. The value of $\bar{r}$ is before normalization.

| Cluster ID | C1n (Red) | C2n (Blue) | C3n (Green) |
|---|---|---|---|
| $S'$ ($\times 10^{-3}$) (%/1000 Å) | $-1.6 \pm 0.5$ | $-2.1 \pm 0.4$ | $-2.0 \pm 0.5$ |
| $D_{550}$ (%) | $2.3 \pm 2.0$ | $1.4 \pm 1.6$ | — |
| Center$_{550}$ (nm) | $578 \pm 14$ | $572 \pm 16$ | — |
| $D_{700}$ (%) | — | — | $0.8 \pm 1.1$ |
| Center$_{700}$ (nm) | — | — | $697 \pm 2$ |
| $\bar{r}$ (550 nm) | $0.042 \pm 0.003$ | $0.044 \pm 0.003$ | $0.046 \pm 0.005$ |
| Area (%) | 26.6 | 52.3 | 21.0 |

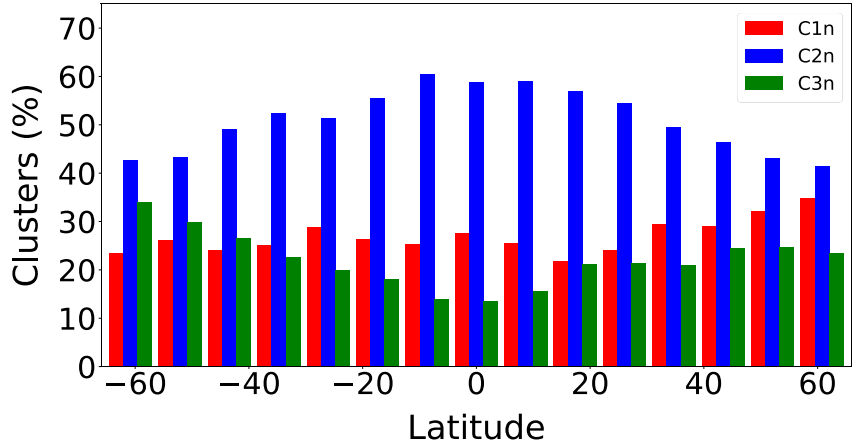

**Figure 5.** Percentage of clusters as a function of the latitude for the normalized global mosaic of Bennu (Figure 3). For this representation we divide the surface into 15 bands (Δ-latitude= 8.68º) and calculate the relative distribution of clusters, avoiding the latitudinal distortion caused by the projection.

### 3.1. High-resolution local images of Nightingale and Osprey

For the primary (Nightingale) and backup (Osprey) sample collection sites of the OSIRIS-REx mission, we analyzed both absolute and normalized reflectance. Absolute reflectance analysis is dominated by differences in reflectance—as shown in our global analysis—and not in spectral shape. Therefore, it only informs the observed albedo range. Normalization involves removing a degree of freedom from our system, and therefore the solution is more compact. Thus, we focus our examination of the primary and backup on spectral clustering using normalized reflectance.

**Nightingale**

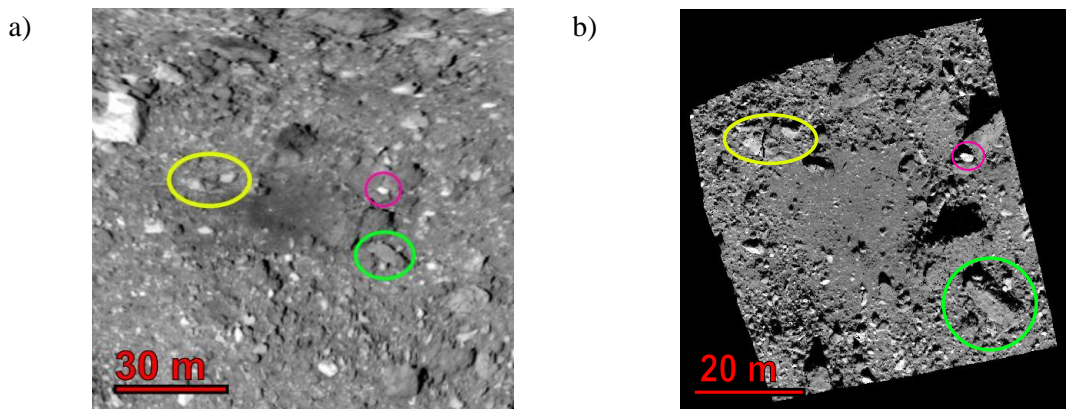

**Figure 6.** Nightingale $b'$ images from (a) Flyby 2B and (b) Recon A. Colored circles point out the same features in both images. Different viewing geometry and pixel scale, and therefore different projection of shadows, can make it difficult to identify the same features in each image.

Nightingale is in a crater of 20 m diameter near Bennu's north pole. We consider Flyby 2B MapCam images acquired at 3.8 km altitude with a pixel scale of 25 cm and Recon A MapCam images acquired at 1.2 km altitude with a pixel scale of 6 cm (Figure 6). Although the higher-resolution Recon A images show more detailed terrain, they were acquired at a higher phase angle (30º versus 8º), which results in greater photometric uncertainty. Photometric corrections shift the measured value of the reflectance to reference angles—in this case $(\alpha, i, e) = (0º, 0º, 0º)$—so accuracy decreases as measured angles tend away from the reference angles. However, both data sets allow us to study how the spectral trends evolve as resolution and phase are increased.

First, we analyze Flyby 2B images (Figure 7a). We detect a spectral behavior at these scales that is similar to what is shown in the global mosaic, which gives us a first hint about the consistency at different pixel scales of MapCam. We found in this case three representative normalized clusters (Figure 7b,c), of which the C1n cluster dominates, concentrated in the darkest areas, confirming the inverse correlation between reflectance and spectral slope (PCC = -0.99 between reflectance at 550 nm and $S_{vwx}$), and also between $b'$ normalized reflectance and slope (PCC = 0.082). In this case, the C1n and C2n clusters present the same UV value, differing only in slope.

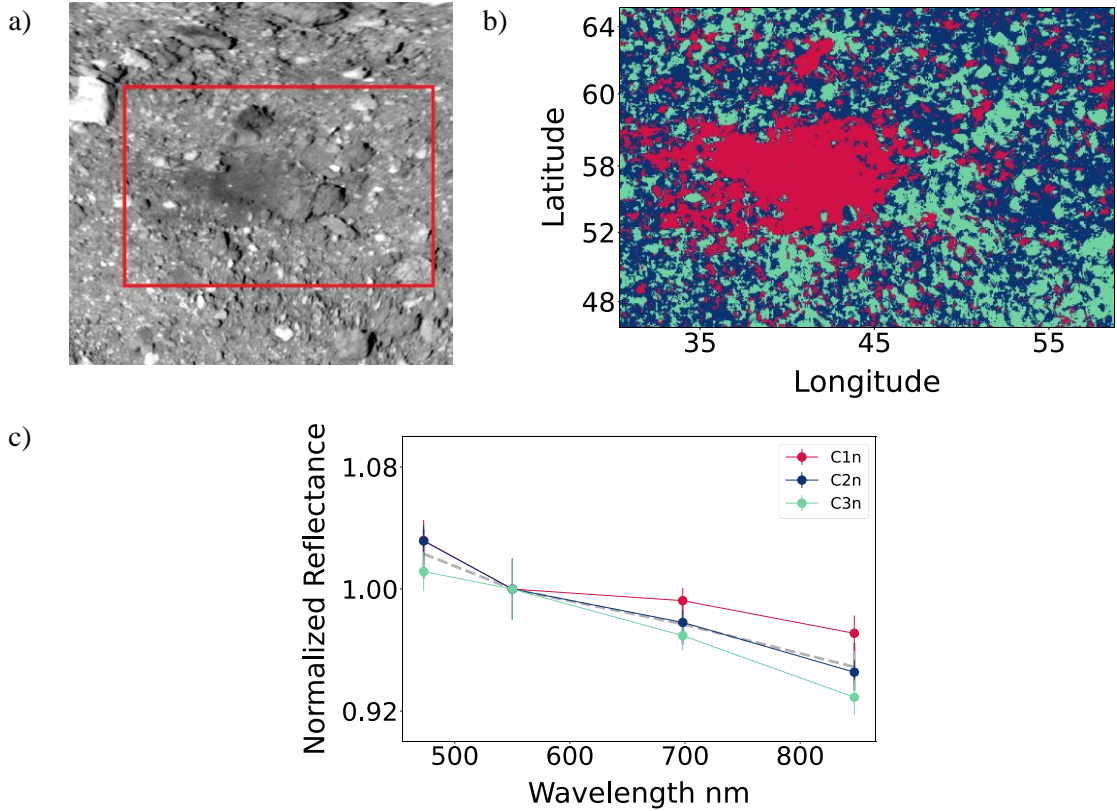

**Figure 7.** Photometrically corrected MapCam image of Nightingale from Flyby 2B ($b'$ filter). The red square encloses the analyzed area. The phase angle is 8º and the pixel scale is 25 cm. b) The location of the identified clusters over the analyzed area. c) The three representative spectral clusters with their corresponding error bars (standard deviation) indicated with different colors, using the same color code as in (b). For comparison, we add the dashed gray line in the background, showing the average normalized Bennu spectrum and its error (standard deviation) obtained from the global mosaic.

Next, we analyze high-resolution Recon A images (Figure 8). In this case much more detail is observed of the terrain, and the C1n cluster is resolved. The three identified clusters show more differences in slope and UV upturn, although errors bars are larger as expected because of the higher phase angle. The distribution of clusters is similar to that observed in the global mosaic: a cluster associated with the darkest areas (C1n), a cluster associated with the intermediate-reflectance areas similar to the ground-based spectra (C2n; highest occupied surface, 44.2%), and a cluster associated with the brightest areas (C3n). The C1n cluster in Nightingale presents the reddest spectra observed overall, -0.3 ± 1.0 × $10^{-3}$ %/1000 Å, and it is the most abundant cluster in the area where the sample was collected (Figure 8a,b). The C1n and C2n clusters allow us to measure UV upturn or absorption band at 550 nm, whereas only the C3n cluster presents a possible absorption feature at 700 nm. However, after our parametric characterization (Table 3), we are not able to identify a clear absorption band either at 550 or 700 nm according to our 3σ criterion.

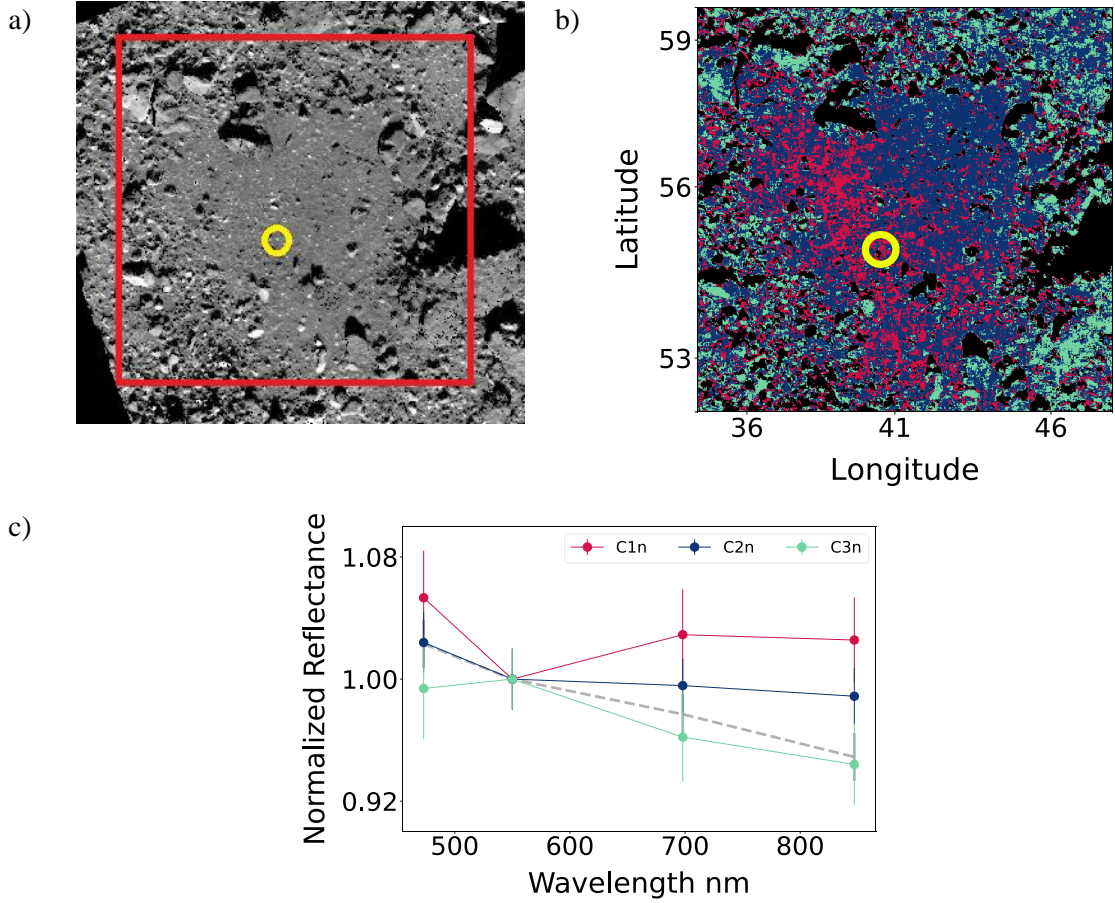

**Figure 8.** Same as Figure 7, but for an image of Nightingale from Recon A. The phase angle is 30º and pixel scale is 6 cm. The yellow circle marks the position where the OSIRIS-REx spacecraft collected a sample on October 20, 2020.

**Table 3.** Same as Table 2, but for the three identified clusters in Nightingale using Recon A data.

| Cluster ID | C1n (Red) | C2n (Blue) | C3n (Green) |
|---|---|---|---|
| $S'$ ($\times 10^{-3}$) (%/1000 Å) | -0.3 ± 1.0 | -0.8 ± 0.7 | -1.5 ± 1.0 |
| $D_{550}$ (%) | 4.8 ± 2.9 | 2.1 ± 2.1 | — |
| Center$_{550}$ (nm) | 582 ± 9 | 576 ± 15 | — |
| $D_{700}$ (%) | — | — | 1.8 ± 2.2 |
| Center$_{700}$ (nm) | — | — | 697 ± 2 |
| $\bar{r}$ (550 nm) | 0.040 ± 0.005 | 0.043 ± 0.005 | 0.048 ± 0.010 |
| Area (%) | 16.4 | 44.2 | 17.0 |

**Osprey**

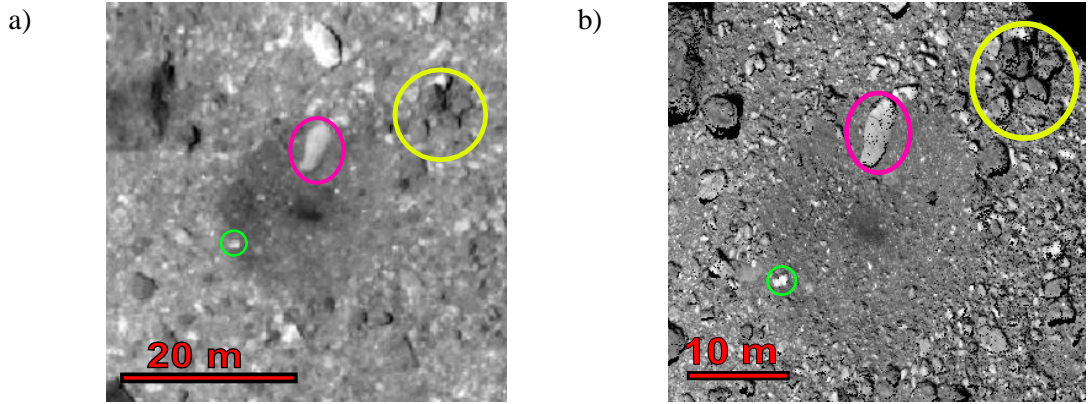

**Figure 9. Osprey *b′* images from (a) Flyby 2B and (b) Recon A, as in Figure 6.**

Osprey is in a crater 20 m diameter near the equator. As we did for Nightingale, we consider MapCam images of Osprey from Flyby 2B with a pixel scale of 25 cm and a phase angle of 8º and MapCam images from Recon A with a pixel scale of 6 cm and a phase angle of 40º (Figure 9).

Our Osprey Flyby 2B analysis identifies three normalized representative clusters without distinction in the UV region and with subtle differences in slope (Figure 10). As noted for Nightingale, the increase in resolution of Recon A images allows us to solve cluster distribution and spectral shape (Figure 11). As with the Nightingale images and global mosaic, the C1n cluster is associated with the darkest terrains, the C3n with the brightest areas, and the C2n (highest occupied area, 45.7%) with intermediate-reflectance areas and resembling ground-based Bennu spectra. After our parametrical characterization (Table 4), we confirm a behavior similar to what we see in the global mosaic, so the homogeneity of the Bennu's surface at different scales from the point of view of MapCam is evident. In this case, the spectral slope of the C1n cluster is lower than Nightingale's: (-0.8 ± 0.8) vs. (-0.3 ± 1.0) × $10^{-3}$ %/1000 Å. A correlation between spectral slope and low reflectance continues to be evident (PCC = -0.99).

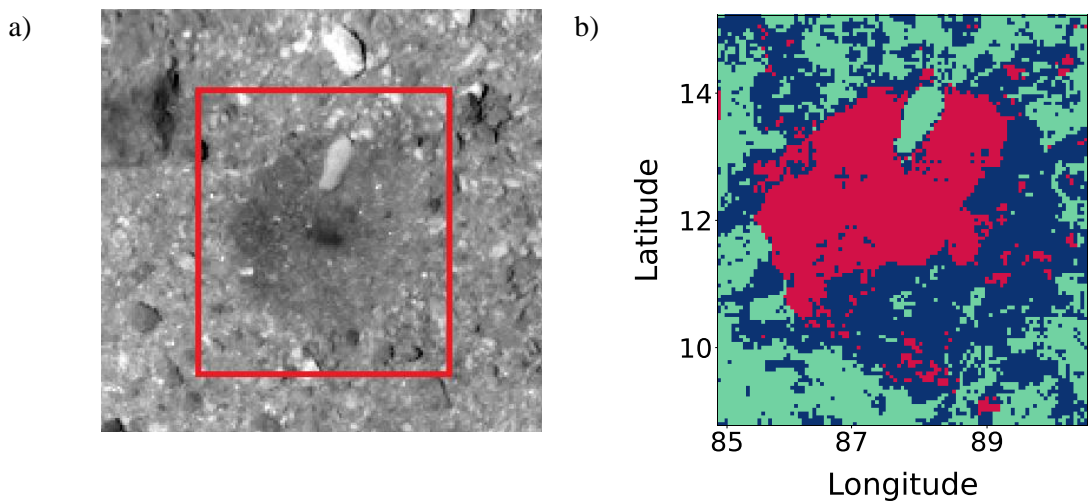

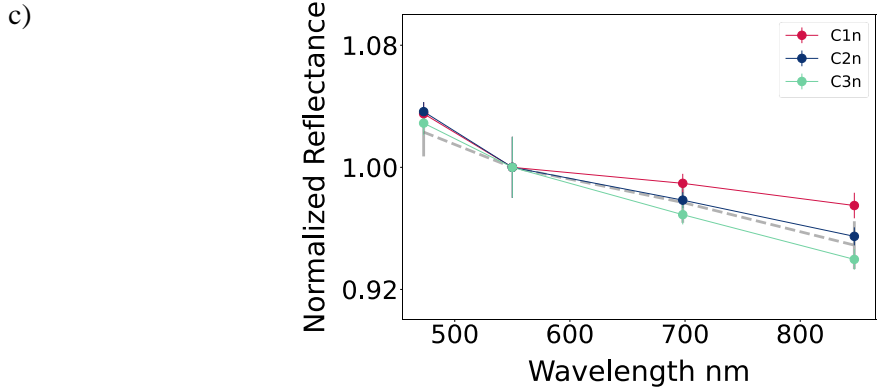

Figure 10. a) Photometrically corrected MapCam image of Osprey from Flyby 2B (*b′* filter). The red square encloses the analyzed area. The phase angle is 8º and pixel scale is 25 cm. b) The location of the identified clusters over the analyzed area. c) The three representative spectral clusters with their corresponding error bars (standard deviation) labeled with different colors, using same color code as in (b). For comparison, we add the dashed gray line in the background, showing the average normalized Bennu spectrum and its error (standard deviation) obtained from the global mosaic.

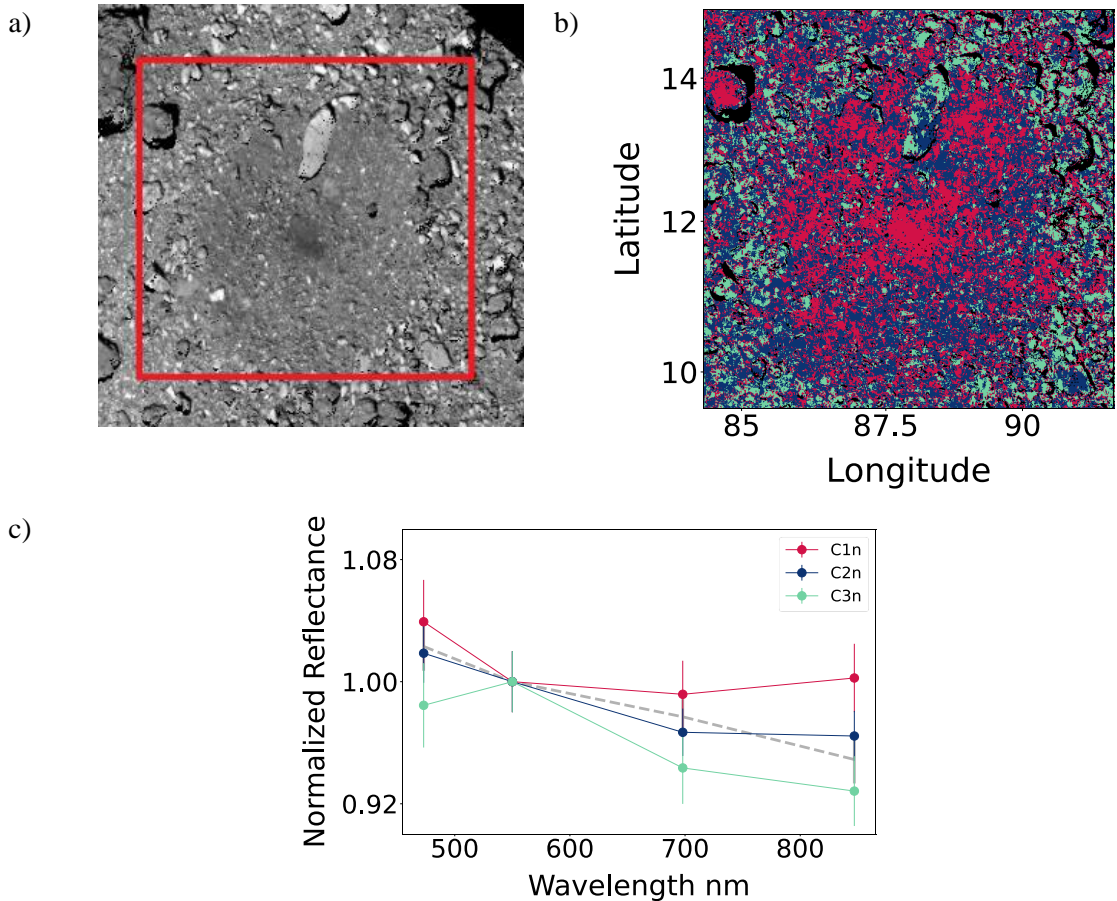

Figure 11. Same as Figure 10, but for an image of Osprey from Recon A. The phase angle is 40º and pixel scale is 6 cm.

After computing absorption band depths, we cannot confirm any absorption band at 550 nm or at 700 nm for Osprey. The higher the phase value, the lower the illumination and, therefore, the SNR. Proof of this lies is in the fact that the absorption band at 550 nm for the C1n cluster (Table 4) is on the same order of the error. On the other hand, we note a different spectral shape from both Nightingale images and the global mosaic. For Osprey, each of the three clusters even shows a small upturn in the *x* filter, making it possible to measure a shallow absorption at 700 nm, although the computed depth does not meet our 3σ criterion. It could also be due to less accurate photometric corrections and, therefore, an incompletely removed phase reddening effect. Note that the phase angle is around 40º, further than the reference phase angle of 0º.

**Table 4. Same as Table 3, but for Osprey.**

| Cluster ID | C1n (Red) | C2n (Blue) | C3n (Green) |
| --- | --- | --- | --- |
| $S'$ ($\times 10^{-3}$) (%/1000 Å) | $-0.8 \pm 0.8$ | $-1.5 \pm 0.6$ | $-1.9 \pm 0.8$ |
| $D_{550}$ (%) | $2.7 \pm 2.4$ | $1.4 \pm 1.6$ | — |
| Center$_{550}$ (nm) | $577 \pm 14$ | $568 \pm 17$ | — |
| $D_{700}$ (%) | $1.6 \pm 1.8$ | $1.8 \pm 1.6$ | $2.5 \pm 2.5$ |
| Center$_{700}$ (nm) | $697 \pm 2$ | $697 \pm 2$ | $697 \pm 2$ |
| $\bar{r}$ (550 nm) | $0.049 \pm 0.003$ | $0.055 \pm 0.008$ | $0.063 \pm 0.015$ |
| Area (%) | 25.9 | 45.7 | 18.9 |

## 4. Discussion and conclusions

The global surface of Bennu is characterized by four clusters scattered across the surface, which reduce to three when we normalize the images at 550 nm. In the absolute reflectance analysis, the four clusters indicate heterogeneity in terms of albedo, corresponding with the bright and dark boulder groups identified in DellaGiustina et al. (2020), who attribute this reflectance dichotomy to primordial heterogeneity. The lowest- and highest-albedo clusters in this study are distributed on rockier terrains. The intermediate-albedo clusters might be the regolith mixture of those two components (Figure 2). Once we normalize the images, our analysis identifies three clusters, also respectively associated with boulders of different reflectance, in which the main difference is in spectral slope and UV value. This result is in agreement with DellaGiustina et al. (2020), where PCA showed the same source of differences.

At the global scale (normalized), we find a cluster occupying ~50% of the surface showing a spectral behavior in agreement with the ground-based Bennu spectra (C2n). The rest of the surface is divided into two clusters: one covering 21% of the surface and presenting both the lowest near UV value and the lowest spectral slope, $(-2.0 \pm 0.5) \times 10^{-3}$ %/1000 Å (C3n); and one covering 26.6% of the surface and presenting a near UV-upturn (or a drop at 550 nm) —although the uncertainty is larger than the depth—, and the most positive spectral slope, $(-1.6 \pm 0.5) \times 10^{-3}$ %/1000 Å (C1n). The C1n cluster corresponds to darker boulders and small craters, whereas the C3n corresponds to the brightest areas distributed over the surface. Both Nightingale and Osprey, which are inside of craters, are classified into C1n clusters within this global analysis. This clustering classification is additionally supported by correlations between normalized *b'* filter reflectance value, spectral slope, and reflectance at 550 nm before normalization. The PCC is -0.99 in the case of reflectance at 550 nm vs. $S_{vwx}$. In the case of *b'* filter normalized reflectance vs $S_{vwx}$, the PCC value is 0.91.

In images of Nightingale and Osprey, we find a similar cluster distribution to that in the global mosaic: three normalized clusters with analogous surface-area proportions. The increased resolution of the Recon A images allows us to find more differences between clusters (both UV and spectral slope), but also leads to larger error bars. In all cases, we find a strong correlation between reflectance and spectral slope, with the darkest areas being the reddest ones. We confirm a correlation between normalized b' filter value and reflectance, where darker areas present a UV upturn or a drop at 550 nm. Nightingale and Osprey have the same spectral behavior in the Recon A images as in the global analysis, so this approach can confirm consistency at different pixel scales. We also find that Nightingale and Osprey are redder than the average Bennu, with Nightingale being the reddest region, with a spectral slope that stands out from the rest: $S' = (-0.3 \pm 1.0) \times 10^{-3}$ %/1000 Å. Although the uncertainties of the slopes are high, both Nightingale and Osprey have spectra redder than 1σ from average; based on the crater chronology by DellaGiustina et al. (2020), this suggests that their craters could have been formed within the past 100,000 years. Given that a redder spectral slope indicates a fresher surface on Bennu, we infer that Nightingale is more recently exposed than Osprey. Moreover, at Nightingale, the C1n-cluster materials are spread continuously from inside to outside of the crater, whereas we do not see the same continuous distribution for Osprey (see Figure 7b and Figure 10b). The C1n-cluster material outside the crater at Nightingale could be an ejecta ray from its crater, indicating recent formation of the crater. Such a morphology can be seen for the artificial crater on asteroid (162173) Ryugu created by the Hayabusa2 mission (Arakawa et al., 2020). The lack of an ejecta ray at Osprey is consistent with an older crater formation age than at Nightingale.

A similar statistical analysis of Bennu was carried out using OVIRS spectra (Barucci et al. 2020). The authors used spectra normalized at 0.55 µm and found four clusters with different spectral slopes. They found that clusters distributed around the equatorial region are redder than the background, i.e., the group containing most of the data and distributed over the entire surface —in opposition to Ryugu, where the equatorial regions are bluer. In our analysis, we find that the C1n cluster presents a slight increase in concentration towards norther regions, although it is not statistically significant. The C2n cluster dominates the equatorial regions with a clear significance. Our bluer cluster, C3n, shows its lowest ratio in equatorial regions. Therefore, we find that our equatorial region does not differ from the background but itself represents the background, mainly due to the fact that there are only 3 clusters characterizing the surface. However, both C1n and C2n are the redder spectra, so is not in conflict with the result shown by Barucci et al. (2020). Our findings are also consistent with theirs in the identification of a correlation between reflectance and spectral slope, as well as a redder spectral slope for Nightingale.

A recent OVIRS study by Li et al. (submitted), also find an equatorial region redder than the global average in the 500 – 2000 nm spectral range. However, they note that if the 440 - 640 nm range is analyzed separately, the equatorial zone is bluer. This result suggests a different spectral trend for that range, which should be also visible in our analysis. And indeed, we also see how the cluster C3n —which is less common in the equatorial region— does not present the near UV-upturn, consistent with the finding that the equatorial region is bluer in this range.

The Hayabusa2 team conducted a global clustering analysis for Ryugu, in this case identifying a non-uniform distribution of clusters (Barucci et al., 2019; Honda et al., 2019). Specifically, they found a latitudinal variation of spectra, which may be caused by the surface processes on Ryugu. We do not see such latitudinal variation on Bennu, which could suggest either better mixed surface materials or very stable surface. Given that geological studies (Jawin et al., 2020; Walsh et al., 2019) and Plastic Finite Element Model (FEM) calculations (Hirabayashi et al., 2020) have indicated high mobility of surface materials on Bennu, we consider the first explanation, a better mixed surface, as the most plausible.

*Spectral features and physical interpretation*

We applied our method to compute absorption features (Section 2.3) considering the error of each band. Our criterion is highly demanding given that we set a minimum value of 3σ to consider a detection as positive.

In all analyses (global and primary and backup sample sites), we see hints of a weak absorption band at 550 nm. The deepest one is found at Nightingale using Recon A images (Table 3), where it reaches a band depth of $(4.8 \pm 2.9)$% for the C1n cluster. This value does not meet our criterion to be considered as a positive detection. Nevertheless, other studies (Hamilton et al., 2019; Lauretta et al., 2019; Simon et al., 2020a; DellaGiustina et al. 2020) attribute this spectral feature to the presence of magnetite, so we must consider this scenario as a real possibility. If we consider that MapCam has only two bands in this spectral region ($b'$ and $v$ filters), another possibility is that we are detecting the presence of a near UV upturn. Indeed, Hendrix et al., (2006, 2019) found evidence of space weathering effects in the UV range on C-type asteroids, so it may explain a near UV-upturn on Bennu. In this case, our latitudinal distribution indicates that the equatorial region presents the most space-weathered materials on Bennu. However, there are not enough studies to support this hypothesis in the case of primitive materials such as carbonaceous chondrites so far. Regardless of the nature of this feature, be it a 550-nm absorption band or a UV upturn, we find a clear correlation between this feature, redder spectral slope, and darker regions, also in agreement with DellaGiustina et al. (2020) and Simon et al. (2020a).

Regarding the other possible absorption band at 700 nm, only in the C3n clusters is possible to measure a (very shallow) band. The deepest absorption is found at Osprey, where it reaches a band depth of (2.5 ± 2.5) %, but it does not exceed 1σ. Therefore, we cannot consider the presence of this band according to our criterion. These values are in agreement with DellaGiustina et al. (2020), where an absorption band was measured in some dark boulders, but at the limit of the relative precision of MapCam. Hamilton et al. (2019) found evidence of abundant hydrated minerals on the surface in the form of a near-infrared absorption near 2.7 μm. The 700-nm absorption band is attributed to an $Fe^{+2} \rightarrow Fe^{+3}$ charge transfer transition in oxidized Fe found in phyllosilicates, so a possible explanation could be either that these hydrated minerals lack oxidized Fe, or that oxidized Fe is present, but the concentration is too low to be detected with MapCam at that SNR.

The spectral slopes have implications for the sample that will be returned, because those belonging to Nightingale would be more pristine and less weathered than Osprey's ones. Nevertheless, other variables may explain such behavior. For example, finer-graine samples tend to show redder spectra (Cloutis et al., 2018). Also, material phase, material maturation, porosity, or roughness could contribute to modify the original spectral behavior. On October 20 2020, OSIRIS-REx successfully collected a sample from the Nightingale sample site, so this conundrum will be addressed once the samples returns to Earth in 2023, allowing extensive laboratory experiments.

## 5. Acknowledgements

We thank the entire OSIRIS-REx Team for making this mission possible. J. L. Rizos, J. Licandro, J. de León and M. Popescu acknowledge support from the AYA2015-67772-R (MINECO, Spain). J. L. Rizos acknowledges Lee Makamson for his invaluable help. J. de León acknowledges financial support from the Severo Ochoa Program SEV-2015-0548 (MINECO) and the project ProID2017010112 under the Operational Programmes of the European Regional Development Fund and the European Social Fund of the Canary Islands (OP-ERDF-ESF), as well as the Canarian Agency for Research, Innovation and Information Society (ACIISI). M. Pajola was supported for this research by the Italian Space Agency (ASI) under the ASI-INAF agreement no. 2017-37-H.0. D.R Golish, H. Campins, D.N. DellaGiustina, and D.S. Lauretta were supported by NASA's OSIRIS-REx mission under Contract NNM10AA11C issued through the New Frontiers Program. This research has made use of the USGS Integrated Software for Imagers and Spectrometers (ISIS).

## 6. Data Availability

OCAMS/MapCam images collected during Detailed Survey and Recon A are available via the Planetary Data System (Rizk et al., 2019). Shape models of Bennu are available via the Small Body Mapping Tool (http://sbmt.jhuapl.edu/).